\documentclass[10pt,conference]{IEEEtran}
\IEEEoverridecommandlockouts
\usepackage{cite}
\usepackage{amsmath,amssymb,amsfonts}
\usepackage{algorithmic}
\usepackage{graphicx}
\usepackage{textcomp}
\usepackage{xcolor}
\usepackage{caption}
\usepackage{listings}
\usepackage{url}
\usepackage{numprint}
\usepackage{flushend}
\def\BibTeX{{\rm B\kern-.05em{\sc i\kern-.025em b}\kern-.08em
    T\kern-.1667em\lower.7ex\hbox{E}\kern-.125emX}}

\newcommand{\KaggleTorrent}{\textsc{KGTorrent}}

\begin{document}

\title{KGTorrent: A Dataset of\\ Python Jupyter Notebooks from Kaggle}

\author{\IEEEauthorblockN{Luigi Quaranta}
\IEEEauthorblockA{\textit{University of Bari, Italy} \\
luigi.quaranta@uniba.it}
\and
\IEEEauthorblockN{Fabio Calefato}
\IEEEauthorblockA{\textit{University of Bari, Italy} \\
fabio.calefato@uniba.it}
\and
\IEEEauthorblockN{Filippo Lanubile}
\IEEEauthorblockA{\textit{University of Bari, Italy} \\
filippo.lanubile@uniba.it}
}

\maketitle

\begin{abstract}
Computational notebooks have become the tool of choice for many data scientists and practitioners for performing analyses and disseminating results. Despite their increasing popularity, 
the research community cannot yet count on a large, curated dataset of computational notebooks. In this paper, we fill this gap by introducing \KaggleTorrent, a dataset of Python Jupyter notebooks with rich metadata retrieved from Kaggle, a platform hosting data science competitions for learners and practitioners with any levels of expertise.
We describe how we built \KaggleTorrent, and provide instructions on how to use it and refresh the collection to keep it up to date. Our vision is that the research community will use \KaggleTorrent{} to study how data scientists, especially practitioners, use Jupyter Notebook in the wild and identify potential shortcomings to inform the design of its future extensions.
\end{abstract}


\section{Introduction}


Computational notebooks, a modern implementation of the literate programming paradigm \cite{knuth_literate_1984}, are interactive documents interleaving natural language text, source code, and its output to form a human-friendly narrative of a computation. The most prominent example of a computational notebook platform is Jupyter Notebook,\footnote{\url{https://jupyter.org}} which has seen a widespread endorsement, especially by data scientists \cite{perkel2018jupyter}. 

Because of their popularity, Jupyter notebooks have also become the primary target of many archival studies \cite{rule_exploration_2018, pimentel_large-scale_2019, koenzen_code_2020, wang_better_2020, yan_auto-suggest_2020}, in which a sizable number of publicly available notebooks from online software repositories are put together under the lens of researchers. However, the task of gathering a large dataset of notebooks, which meets specific research criteria, is nontrivial and time-consuming. 
Due to the novelty of this research area, a large, annotated dataset of computational notebooks has been missing so far.

To fill this gap, in this paper we present \KaggleTorrent{}, a large dataset of computational notebooks with rich metadata retrieved from Kaggle\footnote{\url{www.kaggle.com}}, a Google-owned platform that hosts machine learning competitions for data scientists of all experience levels. In addition to hosting data science challenges, Kaggle also provides a large number of datasets as well as a cloud-based data science environment. The latter enables the development and execution of scripts and computational notebooks written in R or Python. 

Among the various datasets offered by the platform, there is Meta Kaggle,\footnote{\url{https://kaggle.com/kaggle/meta-kaggle}} a daily-updated collection of data about the Kaggle community and its activity. Moreover, Meta Kaggle stores detailed information about publicly available notebooks, which can be  obtained through the Kaggle API or, at a lower level, through direct HTTP requests. 

To build \KaggleTorrent{}, we thoroughly analyzed Meta Kaggle and reverse-engineered its underlying data schema explicit to build a relational database for storing Kaggle metadata; then, after populating our database, we gathered a full copy of \numprint{248761} publicly available Jupyter notebooks written in Python. By linking the notebook archive to the relational database, not only we offer to the research community a large dataset of Jupyter notebooks, but also a practical way to select a sample of interest, based on any criterion that can be expressed in terms of the Kaggle metadata. 


Finally, along with the dataset and its companion database, we publish the scripts used to build them. These can be conveniently executed to reproduce the collection as well as to effortlessly update it to more recent versions of Meta Kaggle. To the best of our knowledge, \KaggleTorrent{} is the largest available dataset of Python Jupyter notebooks with rich metadata.

The name of our dataset is inspired by two previous works of similar nature, GHTorrent \cite{gousios_ghtorrent_2012} and SOTorrent \cite{baltes_sotorrent_2018}. The former provides an offline mirror of data from GitHub, the popular project hosting site. 
The latter is an open dataset containing the version history of posts from Stack Overflow, the most popular question-and-answer website for software developers.


The remainder of this paper is organized as follows. In Section~\ref{datasource}, we present an overview of the Kaggle platform. Next, we describe the two main components of \KaggleTorrent, namely the database of metadata in Section~\ref{sec:database} and the dataset of Jupyter notebooks in Section~\ref{sec:dataset}. Then, Section~\ref{sec:how-to} provides a short guide on how to use and update \KaggleTorrent, while Section~\ref{sec:potential-research-applications}  offers a couple of insights on its potential applications in research. Finally, we describe future work in Section~\ref{sec:conclusions}.

\section{Kaggle}
\label{datasource}

Since 2010, Kaggle started offering worldwide machine learning competitions that ensure their winners both money prizes and high visibility in the platform leaderboards. Notably, some competitions have even been used by international companies for recruiting purposes. Most of the challenges are indeed sponsored by large organizations seeking AI-based innovative approaches to their business challenges or research agenda. Besides providing funds for the competition prizes, they often supply new datasets to the platform, as part of the competition packages. 

Since its foundation, Kaggle has continually evolved over the years, gradually widening its offer to a cloud-based ecosystem of services in support of competitions. It started hosting a large number of datasets, shaping up to be a public data platform, and, most interestingly, began providing its users with a web-based data science environment powered by a state-of-the-art containerization technology. Kaggle enables its users to create scripts and computational notebooks in R and Python -- both known as \textit{`kernels'} in the Kaggle jargon. These can be developed directly on the platform, where large datasets are one click away. The entire computation happens in a containerized cloud environment that users can customize at their will (e.g., by installing custom dependencies). Nevertheless, a comprehensive number of commonly used data science packages are available in kernels out of the box. In addition, both kernels and datasets get versioned in Kaggle. Once a user has finished working on their notebook, they can choose to temporarily save it or commit it -- and, when this applies, submit its results (e.g., a pickled model) in response to a competition. 


Besides competitions and data science tools, Kaggle hosts a rich bundle of social features. The platform enables its users to discuss in forum-like threads about kernels, datasets, and the competitions themselves. 
Additionally, users can follow each other so that content published by the followed user (kernel, datasets, and discussion posts) surfaces in the newsfeed of the follower.

Another key mechanism of the platform is the Kaggle Progression System.\footnote{\url{www.kaggle.com/progression}} The growth of users as data scientists gets tracked in terms of four categories of expertise, namely \textit{Competitions}, \textit{Notebooks}, \textit{Datasets}, and \textit{Discussion}. For each category of expertise, Kaggle assigns its users a performance tier among the following: \textit{Novice}, \textit{Contributor}, \textit{Expert}, \textit{Master}, and \textit{Grandmaster}. Advancement through performance tiers is tied to the activity of users in the platform and abide by different rules depending on the specific category of expertise. 
To step towards the next available tier, users have to earn medals, which are awarded as a consequence of successful action on the site (e.g., a competition won, upvotes received by a forum message, or an uploaded original dataset).


Finally, Kaggle makes freely available \textit{Meta Kaggle}, the official collection of public metadata on users, competitions, datasets, and notebooks from the platform. It was first released on Sept. 2015 and is updated daily, with each new version replacing the previous. 

\section{\KaggleTorrent{} Database}
\label{sec:database}

\KaggleTorrent{} comprises 1) a dataset of Python Jupyter notebooks from Kaggle and 2) a companion database derived from Meta Kaggle, which stores metadata about each notebook and comprehensive information about the overall activity of Kaggle users.\footnote{\KaggleTorrent{} is available at \url{https://doi.org/10.5281/zenodo.4468522}}

\begin{figure*}[tb]
    \centering
    \includegraphics[width=13cm]{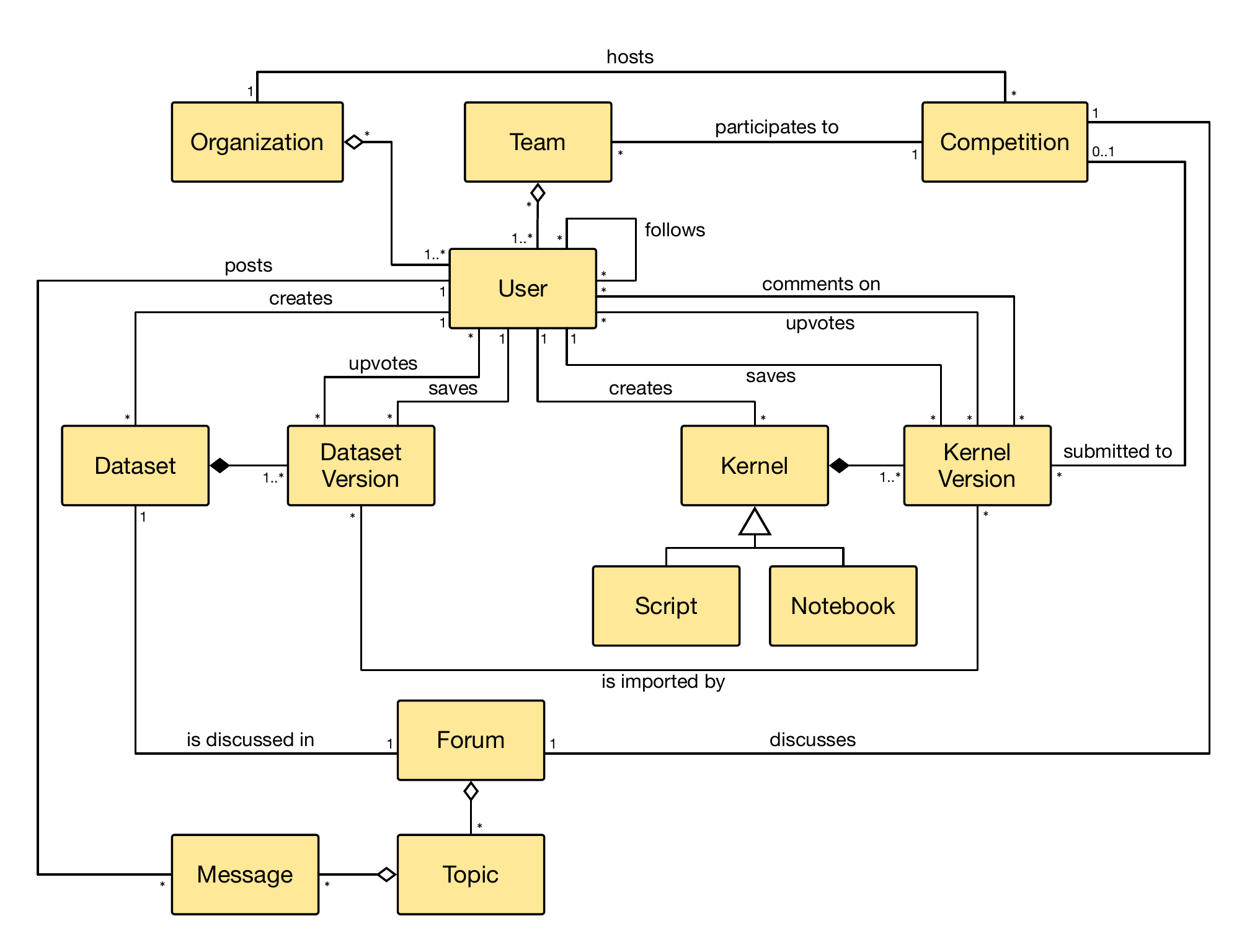}
    \caption{Conceptual schema of the \KaggleTorrent{} database}
    \label{fig:conceptual-schema}
    \vspace{-2mm}
\end{figure*}

As a first step in the development of \KaggleTorrent{}, on October 27, 2020, we downloaded the latest available version of the Meta Kaggle dataset. Meta Kaggle comprises 29 files in the \texttt{.csv} format. Each of these represents the dump of a database table.
Since an official schema definition for Meta Kaggle was not available, we reverse-engineered the relational schema from the set of plain text \texttt{.csv} tables. This step involved understanding the structure of each table, the related constraints as well as the column data types. 
Unfortunately, Kaggle does not provide any official documentation of the dataset, leaving the table content and relationships open to interpretation.
Nevertheless, we managed to piece together the schema structure and column definitions, also by leveraging the related discussions in the forum.

Then, we imported the information contained in Meta Kaggle into a dedicated relational database. Our DBMS of choice was MySQL and that is the format in which we provide the dump of the \KaggleTorrent{} database, weighing 8.31GB (1GB compressed). Nonetheless, users interested in adopting a different DBMS technology can easily create their own version of \KaggleTorrent{}, as we handled all database operations via SQLAlchemy,\footnote{\url{www.sqlalchemy.org}} a popular ORM package for Python supporting a large number of DBMS; therefore, minimal changes are required in our scripts to migrate to a different database technology.

Because some information is missing in the Meta Kaggle dump (i.e., it is not made publicly available by the platform maintainers), many tables present rows for which one or more foreign keys cannot be resolved. Henceforth, a straight import of Meta Kaggle tables using a relational DBMS is not feasible due to a substantial number of violations of referential integrity constraints. To import Meta Kaggle into a relational database, one has to decide whether each foreign key constraint has to be enforced (thus losing orphan records in the referencing table) or dropped (at the expense of the expressiveness of the resulting schema).
We opted for the first solution: to enforce referential integrity, we decided not to insert rows for which one or more foreign keys could not be resolved. While dropping rows from the dataset likely resulted in giving up potentially useful information, we opted for consistency over completeness. In the average case, we kept 91.23\% of the rows of a table -- e.g., for the \textit{Kernels} table, we had to give up 7.69\% of the rows to enforce referential integrity.
Furthermore, we identified several occurrences of cyclic relationships across the tables: to smoothly populate the database and avoid incurring in referential integrity errors during the process, we had to postpone the definition of foreign key constraints after the population step. 

The UML-based conceptual schema in Figure~\ref{fig:conceptual-schema} provides a high-level overview of the database structure and is not meant to capture every detail. 
For instance, it does not show that competitions, kernels, and datasets can be assigned tags.
Similarly, for the sake of brevity, the schema omits relevant entities such as \textit{UserAchievements}, describing the progress of users through the Kaggle Progression System, and the \textit{Submissions} table. For further details, please refer to the logical schema of the database available online as part of the \KaggleTorrent{} documentation.\footnote{\url{https://collab-uniba.github.io/KGTorrent}}

Beyond kernels metadata, the \KaggleTorrent{} database stores information about \numprint{5598921} users and \numprint{2910} competitions.

\section{\KaggleTorrent{} Dataset}
\label{sec:dataset}

The \KaggleTorrent{} dataset consists of \numprint{248761} Jupyter notebooks written in Python. The collection covers the period between November 2015 and October 2020.
The \KaggleTorrent{} dataset has a total size of 175 GB. The compressed notebooks archive has a size of 76 GB.


Kaggle kernels can be downloaded manually via the web interface or through the platform API. 
Interestingly, when downloaded via the API, notebooks lack the output of code cells, thus making impossible to evaluate the results of their execution. 
Being able to examine the output of computations can be crucial to answering certain research questions, such as ``Are computational results replicable?''; therefore, to build \KaggleTorrent{} -- rather than using the Kaggle API -- we pulled integral notebook versions 
via HTTP requests, thus emulating the manual download option.
Unfortunately, we were not able to retrieve every notebook for which we could assemble the download URL as in 9 cases (0.004\%) the HTTP request raised an error.

For each notebook to be downloaded, our scripts build a list of notebook URLs by querying the \KaggleTorrent{} companion database. In particular, the URLs are inferred from the \texttt{CurrentKernelVersionId} field in the \texttt{Kernels} table. We downloaded the latest available version of each kernel. To optimize download duration and resource consumption, we decided to restrict the download to Jupyter notebooks written in Python, as they appear to be the most popular ($\sim$96\% of the total number of kernels). 

\section{Using \KaggleTorrent{}}
\label{sec:how-to}

The most basic use case we envision for \KaggleTorrent{} is the analysis of its whole collection of notebooks. In such a case, downloading the companion database is not required. 

On the other hand, we believe that the most interesting use cases will involve querying the
database for notebooks that meet specific criteria (e.g., high-quality, contest-winning notebooks, notebooks labeled with certain tags, etc.). In such cases, users need to download both the notebooks archive and the companion database dump; then, they should extract the former to a local folder and import the latter in a local installation of MySQL. 

The Jupyter notebooks in \KaggleTorrent{} are saved with a filename based on the following pattern:
\texttt{UserName\_CurrentUrlSlug}.\footnote{Where \texttt{UserName} is a field of the \texttt{Users} table, while \texttt{CurrentUrlSlug} is a field of the \texttt{Kernels} table.}
By including such pattern in the \texttt{SELECT} statement of a database query, the corresponding result set will comprise a column listing the file paths of all the selected notebooks; these will be relative to the position of the folder where users chose to place the uncompressed notebooks archive.

The \KaggleTorrent{} repository\footnote{\url{https://doi.org/10.5281/zenodo.4472989}} contains all the scripts that are necessary to replicate the collection or update it according to a newer version of Meta Kaggle.

\subsection{Replicating the collection}

To replicate the collection, one should run the main script of \KaggleTorrent{} on a Python 3.7+ interpreter with the \texttt{init} argument. Users can choose between one of the two download strategies introduced in Section~\ref{sec:dataset}: the argument \texttt{strategy} accepts either the \texttt{HTTP} option (for a download of full copies of notebooks based on HTTP requests) or the \texttt{API} option (for a faster download of notebooks without code cell outputs, performed via the official Kaggle API). 

Further information needed to run the scripts -- e.g., the details about the connection to the desired MySQL server, or the path to the local folder containing Meta Kaggle -- are expected as environment variables.

\begin{lstlisting}[language=bash, title={Replicate}]
python kgtorrent.py init --strategy HTTP
\end{lstlisting}


If it does not exist yet, the program will create a new MySQL database -- or overwrite the existing one otherwise; then, it will populate the database with Meta Kaggle data, and start the download of
notebooks from Kaggle. Notably, this last step requires a large and increasing amount of time as the number of available Kaggle notebooks keeps growing. As of this writing, we were able to complete the download in about 4.3 days.

\subsection{Refreshing the collection}

Some users might be interested in refreshing \KaggleTorrent{} according to the latest available version of Meta Kaggle. In this case, the \KaggleTorrent{} dataset has to be downloaded and uncompressed along with the new Meta Kaggle version.
Only after performing these preliminary steps and setting the environment variables accordingly, users can run the refresh procedure by issuing the following command:

\begin{lstlisting}[language=bash, title={Refresh}]
python kgtorrent.py refresh --strategy HTTP
\end{lstlisting}

The program will use the latest version of Meta Kaggle to build a fresh copy of the \KaggleTorrent{} database.
Then, the download procedure will start: only the notebooks that are not already present in the local archive will be downloaded and added to the collection; on the other hand, notebooks from the original version of \KaggleTorrent{} that are no more referenced in the refreshed database will be deleted.
Indeed, we observed that, sometimes, new versions of Meta Kaggle do not include all the previously available rows. 
Consequently, these no more referenced notebooks and users become no longer retrievable via database queries due to the constraint-checking procedure that removes tuples with unresolved foreign keys on import.

\section{Potential Research Applications}
\label{sec:potential-research-applications}

Previous studies analyzed computational notebook archives to understand how data scientists use notebooks and how these tools fit in a typical data science workflow \cite{rule_exploration_2018, chattopadhyay2020s}. In addition, as notebooks have been criticized for inducing poor programming practices \cite{grus_i_2018}, some studies have investigated their quality and reproducibility \cite{pimentel_large-scale_2019, koenzen_code_2020, wang_better_2020} and proposed tools to improve them \cite{wang_assessing_2020, wang_restoring_2020, wang_restoring_2021}. The results of these studies confirm that notebooks are often inundated by poor-quality code, ignoring even basic software engineering principles. However, they did not release any corpus of annotated notebooks (e.g., Pimentel et al.~\cite{pimentel_large-scale_2019} shared the replication package scripts and results, but not the dataset); moreover, none of the studies analyzed notebooks from the Kaggle platform.

Researchers interested in this topic might use \KaggleTorrent{} to analyze notebooks produced by data scientists at different levels of expertise, with the goal of identifying usage patterns and common pitfalls in notebook development and inform the design of notebook-specific quality assurance tools. 

Moreover, following the example of Yan et al. \cite{yan_auto-suggest_2020}, notebooks from the \KaggleTorrent{} dataset could be examined with the intent of gaining a deeper understanding of data science pipelines and design recommender systems for assisting practitioners in the field.

\section{Conclusion and Future Improvements}
\label{sec:conclusions}

In this paper, we described \KaggleTorrent{}, a dataset of Python Jupyter notebooks from Kaggle augmented with a companion database containing rich metadata about notebooks and their authors. Our contribution is three-fold: first, we provide an unofficial documentation of the Meta Kaggle dataset and share its logical schema as part of the \KaggleTorrent{} documentation; second, we build a large dataset of Python Jupyter notebooks with rich metadata; third, we share the scripts used for building the \KaggleTorrent{} dataset as well as keeping it up-to-date over time.

Due to the limited resources at our disposal, the current version of the \KaggleTorrent\ dataset is restricted to only Jupyter notebooks written in Python. Still, we understand that the research community might also be interested in examining R notebooks; therefore, we plan to extend future versions of \KaggleTorrent\ to include kernels developed in any programming language.
In addition, as a future extension, we intend to capture information on the historical evolution of notebooks, following the example of GHTorrent \cite{gousios_ghtorrent_2012} and SOTorrent \cite{baltes_sotorrent_2018}.
Finally, we intend to build a minimal RESTful API to enable the download of filtered subsets of notebooks from \KaggleTorrent{} without the need to store the whole collection locally.

\bibliographystyle{IEEEtran}
\bibliography{IEEEabrv, references}

\end{document}